\documentclass[%
 reprint,
 amsmath,amssymb,
 aps,
]{revtex4-2}

\usepackage{graphicx}% Include figure files
\usepackage{dcolumn}% Align table columns on decimal point
\usepackage{bm}% bold math
                        \def\be{\begin{equation}}
                        \def\ee{\end{equation}}
                        \def\ba{\begin{eqnarray}}
                        \def\ea{\end{eqnarray}}
                        \def\bas{\begin{eqnarray*}}
                        \def\eas{\end{eqnarray*}}

\begin{document}

\title{Analytical solutions of the Schrödinger equation for the one-dimensional hydrogen molecular ion}

\author{Stavros Theodorakis}
                        \affiliation{Physics Department, University of Cyprus,
P.O. Box 20537, Nicosia 1678, Cyprus}
                        \email{stavrost@ucy.ac.cy}
\date{\today}

\begin{abstract}
We present analytical solutions of the Schrödinger equation for the one-dimensional hydrogen molecular ion. In particular, we present closed form expressions for the electronic energy curves of this system that correspond to the ground state and the first excited state. Our results agree with numerical solutions obtained before.  
\end{abstract}

\maketitle

\vskip 0.3cm
\vskip 0.3cm

{\bf I. Introduction}

The one-dimensional hydrogen molecular ion is the simplest two-centered quantum system. Yet, a closed form analytical solution of the corresponding Schrödinger equation has never been presented till now. The only attempt at an analytical solution  involved a series expansion of the wavefunction and the use of continued fractions in order to determine numerically the coefficients of the series and of the energies, for each value of the distance between the protons\cite{Duan}. In a recent work, the appropriate one-dimensional Schrödinger equation was solved numerically, using the fixed-step-fourth-order Runge-Kutta method\cite{Alejandro}. Thus the electronic energy curves were obtained numerically, if the distance between the protons was given. Such purely numerical solutions had been also obtained earlier\cite{Loos}.

In this paper we present analytical solutions of the Schrödinger equation for the ground state and the first excited state of the one-dimensional hydrogen molecular ion. In particular, we present closed form expressions for the electronic energy curves of this system. Our results agree with the numerical solutions obtained before\cite{Duan, Alejandro, Loos}.

The one-dimensional hydrogen molecular ion consists of one electron restricted to move between two stationary protons along the x-axis. The two protons act as impenetrable walls\cite{Nunez, Oseguera}. The Schrödinger equation for this molecular ion is

\be
\label{Schrodinger}
-\frac{\hbar^{2}}{2m}\frac{\partial^{2}\Psi}{\partial x^{2}}-\frac{e^{2}\Psi}{x}-\frac{e^{2}\Psi}{L-x}+\frac{e^{2}\Psi}{L}=E\Psi,
\ee
where the protons are located at $x=0$ and $x=L$. The wavefunction is zero at the locations of the protons. It is also zero for $x\leq 0$ and $L\leq x$. We note that if $L\rightarrow\infty$ then the system reduces to the one-dimensional hydrogen atom\cite{Oseguera}, because the electron remains close to one of the protons and does not feel at all the influence of the other proton. In that case the wavefunction and the energy for the ground state are $xe^{-x/a}$ and $-me^{4}/(2\hbar^{2})$ respectively, where $a=\hbar^{2}/(me^{2})$.

We note furthermore that the attractive part of the potential is $-e^{2}/(L\xi)$, where

\be
\label{xi}
\xi=\frac{x(L-x)}{L^{2}}.
\ee

The variable $\xi$ is an even function about the midpoint $x=L/2$. It is zero on both protons and its maximum value $1/4$ is attained at the midpoint $x=L/2$. We expect the ground state wavefunction and the first excited state wavefunction to be even and odd respectively with respect to this midpoint. Hence we need only examine what happens in the interval $0\leq x\leq L/2$, where

\be
\label{xison}
x=\frac{L}{2}(1-\sqrt{1-4\xi}).
\ee

We shall treat $\Psi$ as a function of $\xi$. Indeed, we define $\Psi=\psi(\xi)/\sqrt{L}$. Then both boundary conditions on the protons reduce to $\psi(0)=0$. We shall also define the dimensionless energy $\epsilon=E\hbar^{2}/(me^{4})$ and the dimensionless distance $\nu=L/a$. Then Eq.~(\ref{Schrodinger}) takes the dimensionless form

\be
\label{dimlessSchrodinger}
\frac{4\xi-1}{2\nu^{2}}\frac{d^{2}\psi}{d\xi^{2}}+\frac{1}{\nu^{2}}\frac{d\psi}{d\xi}+(1-\frac{1}{\xi})\frac{\psi}{\nu}=\epsilon\psi.
\ee

We multiply this equation with $\psi(\xi)/\sqrt{1-4\xi}$ and we integrate from $\xi=0$ to $\xi=1/4$. We obtain then an expression for the dimensionless energy, assuming the wavefunction $\psi$ is normalized:

\ba
\label{energy}
&&\int_{0}^{1/4}\frac{\sqrt{1-4\xi} }{\nu^{2}}\bigl(\frac{d\psi}{d\xi}\bigr)^{2}d\xi\nonumber\\
&&+\frac{1}{\nu}-\frac{2}{\nu}\int_{0}^{1/4}\frac{\psi(\xi)^{2}}{\xi\sqrt{1-4\xi}}d\xi
\ea

The normalization condition $\int_{0}^{L}\Psi(x)^{2}dx=1$ becomes

\be
\label{normalization}
\int_{0}^{1/4}\frac{\psi(\xi)^{2}}{\sqrt{1-4\xi}}d\xi=\frac{1}{2}.
\ee

{\bf II. The ground state}

The ground state wavefunction is even with respect to the midpoint $x=L/2$, just like $\xi$. We shall define then the wavefunction for this even state as

\be
\label{ansatzeven}
\psi(\xi)=N\xi e^{-\nu\xi}f(\xi),
\ee
where $f(0)=1$ and N will be determined by the normalization condition Eq.~(\ref{normalization}). If $L\rightarrow\infty$, then $\xi\approx x/L$ and $x/a\approx L\xi/a=\nu\xi$. Thus in the limit of the one-dimensional hydrogen $f(\xi)=1$ and we obtain $\psi=xe^{-x/a}$, as we must.

We insert Eq.~(\ref{ansatzeven}) into Eq.~(\ref{dimlessSchrodinger}) and we obtain thus the differential equation that $f$ must satisfy:

\ba
\label{eqforf}
&&\frac{\xi}{2}(4\xi-1)f^{\prime\prime}(\xi)+(5\xi+\nu\xi-4\nu\xi^{2}-1)f^{\prime}(\xi)\nonumber\\
&&+(1-4\nu\xi-\frac{\nu^{2}\xi}{2}-\epsilon\nu^{2}\xi+2\nu^{2}\xi^{2})f(\xi)=0
\ea

Since $0\leq\xi\leq 1/4$, a series solution of this equation will converge rapidly. We can easily find then that

\be
\label{fexpansion}
f(\xi)=\sum_{i=0}^{\infty}w_{i}\xi^{i},
\ee

where

\ba
\label{eqw}
&&w_{0}=1\nonumber\\
&&w_{1}=1\nonumber\\
&&w_{2}=\frac{1}{6}(12-6\nu-\nu^{2}-2\epsilon\nu^{2})\nonumber\\
&&w_{3}=\frac{1}{18}(90-57\nu-9\nu^{2}-18\epsilon\nu^{2}-\nu^{3}-2\epsilon\nu^{3})\nonumber\\
&&w_{4}=\frac{1}{360}(5040-3516\nu-378\nu^{2}-1080\epsilon\nu^{2}-20\nu^{3}\nonumber\\
&&-40\epsilon\nu^{3}-3\nu^{4}+12\epsilon^{2}\nu^{4})\nonumber\\
&&w_{5}=\frac{1}{5400}(226800-166860\nu-12294\nu^{2}-50400\epsilon\nu^{2}\nonumber\\
&&+318\nu^{3}+780\epsilon\nu^{3}+75\nu^{4}+600\epsilon\nu^{4}+900\epsilon^{2}\nu^{4}\nonumber\\
&&-2\nu^{5}+40\epsilon\nu^{5}+88\epsilon^{2}\nu^{5})\nonumber\\
&&etc
\ea

Having found the exact series for $f(\xi)$, we can use Eq.~(\ref{normalization}) in order to find N. Then the energy functional of Eq.~(\ref{energy}) can be calculated as a function of the parameters $\nu$ and $\epsilon$. Requiring that the energy calculated be equal to the parameter $\epsilon$ would determine the value of $\epsilon$ in terms of $\nu$ for the various even states. Each root $\epsilon$ of the resulting equation corresponds to an even state. The lowest root corresponds to the ground state.

There is an alternate way though to locate the value of $\epsilon$ that corresponds to the ground state. We shall consider the energy functional that is calculated using Eq.~(\ref{fexpansion}) as a function of a variational parameter $\epsilon$ for a given value of $\nu$. Minimization of this function with respect to $\epsilon$ will yield the exact value of the lowest energy in terms of $\nu$, since the correct wavefunction of Eq.~(\ref{fexpansion}) has been used.

However, since the series is expected to converge rapidly, we can truncate the series at some point we choose. We can then determine N by putting the truncated wavefunction into Eq.~(\ref{normalization}) and use Eq.~(\ref{energy}) to calculate the energy. Our result will involve the parameters $\epsilon$ and $\nu$. However, if the full expression of Eq.~(\ref{fexpansion}) had been used our result would have been exact and the energy functional of Eq.~(\ref{energy}) would have its minimum at the correct value of $\epsilon$. Conversely, we can minimize the energy functional that arose through the use of the truncated $f(\xi)$, with respect to $\epsilon$. The function of $\nu$ we shall find will be an excellent approximation to the total energy.

We shall demonstrate below how rapidly this procedure converges and how we can obtain a very accurate expression for the energy as a function of the dimensionless distance $L/a=\nu$ between the protons.

The simplest choice for a truncated $f(\xi)$ is $1+w_{1}\xi$. Thus our wavefunction for this choice is

\be
\label{wavefunction1}
\psi(\xi)=N_{1}\xi e^{-\nu\xi}(1+\xi)
\ee
We can use Eq.~(\ref{normalization}) to find $N_{1}$ and then we can calculate the energy of Eq.~(\ref{energy}), obtaining a result that contains only the parameter $\nu$:

\be
\label{energy1}
\epsilon_{1}=A_{1}/B_{1},
\ee

where
\ba
\label{numer1}
&&A_{1}=
\sqrt{\nu}(15+7\nu+147\nu^{2}+175\nu^{3})\nonumber\\
&&-\sqrt{2}F(\frac{\sqrt{\nu}}{\sqrt{2}})(15+12\nu+150\nu^{2}+100\nu^{3}+175\nu^{4})\nonumber
\ea

and

\ba
\label{denomin1}
&&B_{1}=-10\nu^{3/2}(21+29\nu+17\nu^{2}+5\nu^{3})\nonumber\\
&&+2\sqrt{2}\nu F(\frac{\sqrt{\nu}}{\sqrt{2}})(105+180\nu+138\nu^{2}+60\nu^{3}+25\nu^{4}),\nonumber
\ea
where $F(x)=e^{-x^{2}}\int_{0}^{x}e^{y^{2}}dy$ is the Dawson integral.

\begin{figure}[t]
\vskip 0.3cm
                        \includegraphics[width=0.49\textwidth]{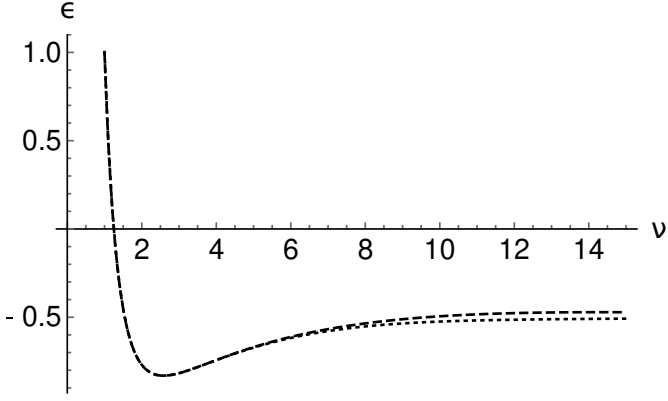}
                        \caption{\label{fig1}The dimensionless ground state energy $\epsilon$ as a function of $\nu=L/a$ for the truncated wavefunction of Eq.~(\ref{wavefunction1}). The dashed curve is the analytical result of Eq.~(\ref{energy1}), while the dotted curve is the numerical result of Ref.\cite{Duan}.}
                        \end{figure}

For small $\nu$ $\epsilon_{1}\approx 153/(31\nu^{2})$, while for large $\nu$ $\epsilon_{1}\approx (-1/2)+9/(2\nu^{2})$. Clearly, the particular truncated wavefunction we have used so far is inadequate for large $\nu$, because in that region the energy of the molecular ion is greater than the energy of the hydrogen atom. We note that $\epsilon_{1}$ acquires its minimum value of -0.830708 at $\nu=2.58134$, while the numerical result\cite{Loos} for this equilibrium position is $\nu=2.581$ with a corresponding energy of -0.830710. When $\nu=2.6$ it acquires the value -0.830671, while the corresponding numerical result is -0.830672772\cite{Duan}. We clearly see in Figure~\ref{fig1} that the agreement of the analytical expression $\epsilon_{1}$ of Eq.~(\ref{energy1}) with the numerical results of Ref.\cite{Duan} is excellent for $\nu\leq 8$, even though the truncated function is the simplest one we could choose. A clear deviation exists however for $\nu\geq 8$. For example, Eq.~(\ref{energy1}) gives an energy of $-0.494552$ if $\nu=10$, when the corresponding exact numerical value from Ref.\cite{Duan} is $-0.524383$.

												The next simplest choice for a truncated $f(\xi)$ is $1+w_{1}\xi+w_{2}\xi^{2}$. Thus our wavefunction for this choice is

\be
\label{wavefunction2}
\psi(\xi)=N_{2}\xi e^{-\nu\xi}(1+\xi+w_{2}\xi^{2})
\ee
We can use Eq.~(\ref{normalization}) to find:

\be
\label{normalization2}
N_{2}=(k_{0}+k_{1}\epsilon+k_{2}\epsilon^{2})^{-1/2},
\ee
where

\ba
\label{k0}
&&k_{0}=\frac{287}{24576}-\frac{10395}{1024\nu^{6}}+\frac{2205}{512\nu^{5}}-\frac{357}{4096\nu^{4}}-\frac{283}{1024\nu^{3}}\nonumber\\
&&-\frac{3175}{16384\nu^{2}}-\frac{1077}{16384\nu}+\frac{5\nu}{8192}-\frac{19\nu^{2}}{147456}-\frac{\nu^{3}}{147456}\nonumber\\
&&+\frac{F(\sqrt{\nu/2})}{\sqrt{2\nu}}\bigl(\frac{1259}{8192}+\frac{10395}{512\nu^{6}}-\frac{945}{512\nu^{5}}-\frac{3675}{2048\nu^{4}}+\frac{555}{2048\nu^{3}}\nonumber\\
&&+\frac{4575}{8192\nu^{2}}+\frac{1173}{4096\nu}-\frac{35\nu}{1536}-\frac{37\nu^{2}}{24576}\nonumber\\
&&+\frac{\nu^{3}}{4096}+\frac{\nu^{4}}{73728}\bigr),
\ea

\ba
\label{k1}
&&k_{1}=\frac{1}{6144}+\frac{3465}{1024\nu^{4}}-\frac{735}{2048\nu^{3}}-\frac{441}{4096\nu^{2}}-\frac{57}{4096\nu}\nonumber\\
&&+\frac{\nu}{1024}-\frac{13\nu^{2}}{36864}-\frac{\nu^3}{36864}\nonumber\\
&&+\frac{F(\sqrt{\nu/2})}{\sqrt{2\nu}}\bigl(\frac{43}{2048}-\frac{3465}{512\nu^{4}}-\frac{1575}{1024\nu^{3}}+\frac{315}{2048\nu^{2}}\nonumber\\
&&+\frac{15}{128\nu}-\frac{\nu}{3072}-\frac{17\nu^{2}}{6144}+\frac{\nu^{3}}{1536}+\frac{\nu^{4}}{18432}\bigr)
\ea

and
\ba
\label{k2}
&&k_{2}=-\frac{10395+2205\nu+378\nu^{2}+54\nu^{3}+7\nu^{4}+\nu^{5}}{36864\nu^{2}}\nonumber\\
&&+\frac{F(\sqrt{\nu/2})}{\sqrt{2\nu}}\frac{10395+5670\nu+1575\nu^{2}}{18432\nu^{2}}\nonumber\\
&&+\frac{F(\sqrt{\nu/2})}{\sqrt{2\nu}}\frac{300\nu^{3}+45\nu^{4}+6\nu^{5}+\nu^{6}}{18432\nu^{2}}.
\ea

We can also calculate the energy of Eq.~(\ref{energy}), to obtain:

\be
\label{energy2}
\epsilon_{2}=\frac{1}{\nu}+N_{2}^{2}(g_{0}+g_{1}\epsilon+g_{2}\epsilon^{2})
\ee

where
\ba
\label{g0}
&&g_{0}=-\frac{613}{147456}+\frac{21735}{2048\nu^{7}}-\frac{2835}{1024\nu^{6}}-\frac{18759}{
8192\nu^5}+\frac{627}{512\nu^{4}}\nonumber\\
&&+\frac{17939}{32768\nu^{3}}+\frac{37271}{98304\nu^{2}}-\frac{2567}{49152\nu}+\frac{151\nu}{294912}+\frac{\nu^2}{32768}\nonumber\\
&&+\frac{F(\sqrt{\nu/2})}{\sqrt{2\nu}}\bigl(\frac{607}{6144}-\frac{21735}{1024\nu^{7}}-\frac{1575}{1024\nu^{6}}+\frac{22455}{4096\nu^{5}}-\frac{2955}{4096\nu^{4}}\nonumber\\
&&-\frac{27739}{16384\nu^{3}}-\frac{5371}{8192\nu^{2}}-\frac{42497}{49152\nu}+\frac{155\nu}{16384}\nonumber\\
&&-\frac{71\nu^{2}}{73728}-\frac{\nu^3}{16384}\bigr),
\ea
\ba
\label{g1}
&&g_{1}=-\frac{155}{18432}-\frac{7245}{2048\nu^{5}}-\frac{1365}{4096\nu^{4}}+\frac{3973}{8192}{\nu^{3}}\nonumber\\
&&+\frac{595}{24576\nu^{2}}-\frac{65}{12288\nu}+\frac{97\nu}{73728}+\frac{\nu^{2}}{8192}\nonumber\\
&&+\frac{F(\sqrt{\nu/2})}{\sqrt{2\nu}}\bigl(\frac{3}{2048}+\frac{7245}{1024\nu^{5}}+\frac{6195}{2048\nu^{4}}-\frac{1775}{4096\nu^{3}}\nonumber\\
&&-\frac{335}{1024\nu^{2}}-\frac{593}{12288\nu}+\frac{245\nu}{12288}-\frac{11\nu^{2}}{4608}-\frac{\nu^3}{4096}\bigr)
\ea
and

\ba
\label{g2}
&&g_{2}=\frac{155}{36864}+\frac{2415}{8192\nu^{3}}+\frac{1085}{8192\nu^{2}}\nonumber\\
&&+\frac{319}{12288\nu}+\frac{43\nu}{73728}+\frac{\nu^2}{8192}\nonumber\\
&&+\frac{F(\sqrt{\nu/2})}{\sqrt{2\nu}}\bigl(-\frac{119}{3072}-\frac{2415}{4096\nu^{3}}-\frac{945}{2048\nu^{2}}\nonumber\\
&&-\frac{2045}{12288\nu}-\frac{83\nu}{12288}-\frac{17\nu^{2}}{18432}-\frac{\nu^{3}}{4096}\bigr).
\ea

In all the above $F(x)=e^{-x^{2}}\int_{0}^{x}e^{y^{2}}dy$ is the Dawson integral.

The energy of Eq.~(\ref{energy2}) acquires its extrema at two values of $\epsilon$:

\ba
\label{plusminus}
&&\pm\frac{\sqrt{(g_{2}k_{0}-g_{0}k_{2})^{2}+(g_{1}k_{0}-g_{0}k_{1})(-g_{2}k_{1}+g_{1}k_{2})}}{g_{1}k_{2}-g_{2}k_{1}}\nonumber\\
&&+\frac{g_{2}k_{0}-g_{0}k_{2}}{g_{1}k_{2}-g_{2}k_{1}}.
\ea
However, we can easily ascertain that the upper sign corresponds to a maximum for all values of $\nu$. Thus we have to let

\ba
\label{extremum}
&&\epsilon=-\frac{\sqrt{(g_{2}k_{0}-g_{0}k_{2})^{2}+(g_{1}k_{0}-g_{0}k_{1})(-g_{2}k_{1}+g_{1}k_{2})}}{g_{1}k_{2}-g_{2}k_{1}}\nonumber\\
&&+\frac{g_{2}k_{0}-g_{0}k_{2}}{g_{1}k_{2}-g_{2}k_{1}}
\ea

If we insert this value of the variational parameter $\epsilon$ into the energy of Eq.~(\ref{energy2}), we obtain the final result for the energy:

\ba
\label{energy2final}
&&\epsilon_{2}=\frac{g_{1}k_{1}-2g_{0}k_{2}-2g_{2}k_{0}}{k_{1}^{2}-4k_{0}k_{2}}\nonumber\\
&&+\frac{2\sqrt{(g_{2}k_{0}-g_{0}k_{2})^{2}+(g_{1}k_{0}-g_{0}k_{1})(-g_{2}k_{1}+g_{1}k_{2})}}{k_{1}^{2}-4k_{0}k_{2}}\nonumber\\
&&+\frac{1}{\nu}.
\ea

\begin{figure}[t]
\vskip 0.3cm
                        \includegraphics[width=0.49\textwidth]{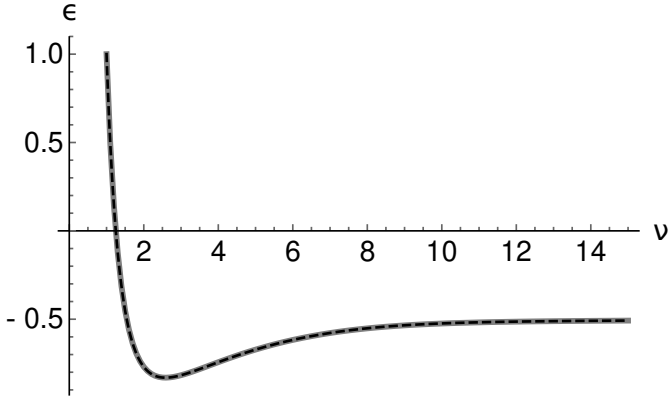}
                        \caption{\label{fig2}The dimensionless ground state energy $\epsilon$ as a function of $\nu=L/a$ for the truncated wavefunction of Eq.~(\ref{wavefunction2}). The dashed curve is the analytical result of Eq.~(\ref{energy2final}), while the continuous gray curve is the numerical result of Ref.\cite{Duan}.}
                        \end{figure}

For small $\nu$ we find that $\epsilon_{2}\approx (6270-\sqrt{25314630})/(251\nu^{2})\approx 4.9348/\nu^{2}$, while for large $\nu$ $\epsilon_{2}\approx (-1/2)-3/(2\nu^{2})$. We note that $\epsilon_{2}$ acquires its minimum value of -0.830709 at $\nu=2.58117$, while the numerical result\cite{Loos} for this equilibrium position is $\nu=2.581$ with a corresponding energy of -0.830710. When $\nu=2.6$ it acquires the value -0.830672, while the corresponding numerical result is -0.830672772\cite{Duan}. When $\nu=10$ it acquires the value -0.523867, while the corresponding numerical result is -0.524383\cite{Duan}. We clearly see in Figure~\ref{fig2} that the agreement of the analytical expression $\epsilon_{2}$ of Eq.~(\ref{energy2final}) with the numerical results of Ref.\cite{Duan} is excellent for all values of $\nu$, even though the truncated function is very simple.

\begin{figure}[t]
\vskip 0.3cm
                        \includegraphics[width=0.49\textwidth]{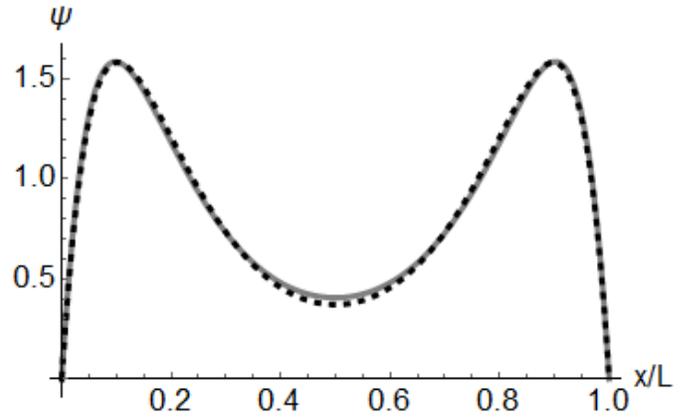}
                        \caption{\label{fig3} The ground state wavefunction of Eq.~(\ref{wavefunction2}) (continuous gray curve) is compared to the corresponding wavefunction obtained numerically for the case $\nu=10$ (dashed curve).}
                        \end{figure}

\begin{figure}[t]
\vskip 0.3cm
                        \includegraphics[width=0.49\textwidth]{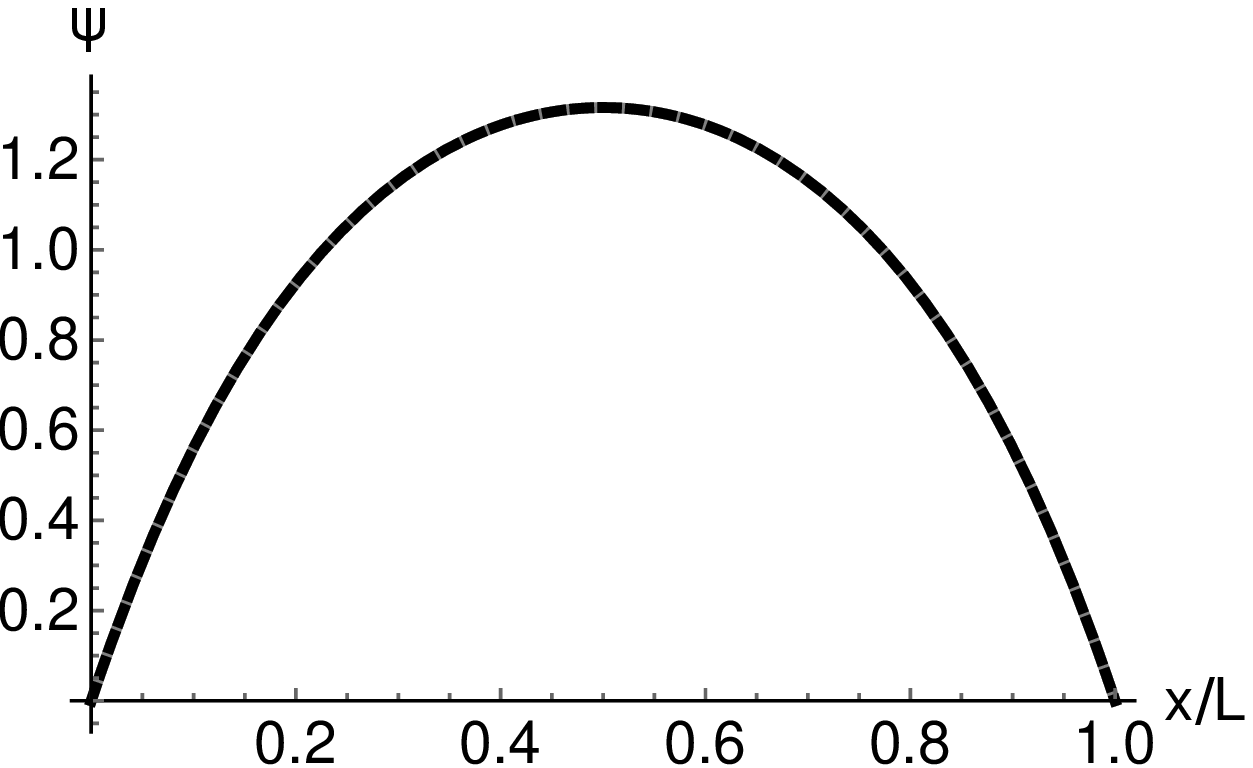}
                        \caption{\label{fig4} The ground state wavefunction of Eq.~(\ref{wavefunction2}) (continuous gray curve) is compared to the corresponding wavefunction obtained numerically for the case $\nu=2$ (dashed curve).}
                        \end{figure}

We can also compare the wavefunction of Eq.~(\ref{wavefunction2}) with the wavefunction obtained numerically\cite{Alejandro}. This comparison is shown in Figure~\ref{fig3} for $\nu=10$ and in Figure~\ref{fig4} for $\nu=2$. We see again an excellent agreement between our analytical results and their numerical counterparts.

\begin{figure}[t]
\vskip 0.3cm
                        \includegraphics[width=0.49\textwidth]{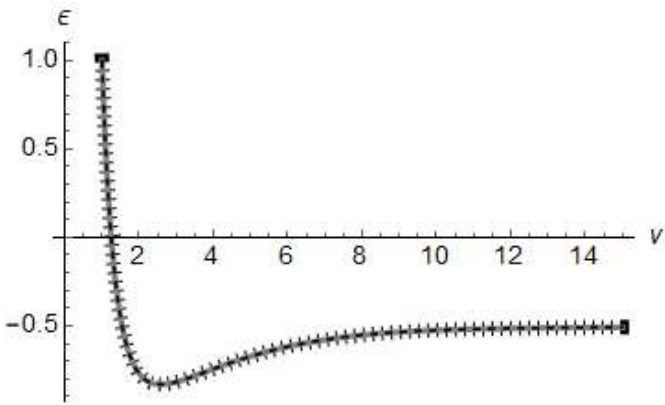}
                        \caption{\label{fig5}The dimensionless ground state energy $\epsilon$ as a function of $\nu=L/a$ for the truncated wavefunction of Eq.~(\ref{wavefunction2}) and for truncated wavefunctions ending with the $w_{3}\xi^{3}$ and $w_{4}\xi^{4}$ terms. The dashed curve corresponds to the truncated wavefunction that ends with the $w_{2}\xi^{2}$ term, the continuous gray curve corresponds to the truncated wavefunction that ends with the $w_{3}\xi^{3}$ term, and the dotted curve corresponds to the truncated wavefunction that ends with the $w_{4}\xi^{4}$ term. All three curves coincide.}
                        \end{figure}

Actually, we can check the accuracy of our analytical results by examining the possibility of adopting as $f(\xi)$ the expression of Eq.~(\ref{wavefunction2}), augmented first by a $w_{3}\xi^{3}$ term and then by an additional $w_{4}\xi^{4}$ term. We find that the results obtained by using Eq.~(\ref{wavefunction2}) as is are identical to the ones obtained when we use the augmentations mentioned above. This is most clearly seen in Figure~\ref{fig5}, where the dashed curve corresponds to the truncated wavefunction that ends with the $w_{2}\xi^{2}$ term, the continuous gray curve corresponds to the truncated wavefunction that ends with the $w_{3}\xi^{3}$ term, and the dotted curve corresponds to the truncated wavefunction that ends with the $w_{4}\xi^{4}$ term. All three curves coincide, showing thus that we need go no further than the wavefunction of Eq.~(\ref{wavefunction2}) in order to describe accurately the ground state of the one-dimensional molecular ion.

\vskip 0.3cm
\vskip 0.3cm

{III. \bf The first excited state}

The wavefunction for the first excited state is odd with respect to the midpoint $x=L/2$. We know that if an even potential consists of two regions of high localization that are distant from each other, then the probability distribution, and hence the wavefunction, will also consist of two regions of high localization. There are however two possibilities that can lead to such a doubly localized probability: an even wavefunction, where the localization in both regions has the same sign, and an odd wavefunction, where the two localizations have opposite sign. The even wavefunction will have a slightly lower energy than the odd one. In the limit that the localizations are infinitely apart, however, these states will be degenerate.

We shall define then the wavefunction for the odd state as

\be
\label{ansatzodd}
\psi(\xi)=N\xi\sqrt{1-4\xi}e^{-\nu\xi}f(\xi),
\ee
where $f(0)=1$ and N will be determined by the normalization Eq.~(\ref{normalization}). This wavefunction becomes zero at the points $x=0,L/2,L$. If $L\rightarrow\infty$, then $\xi\approx x/L$ and $x/a\approx L\xi/a=\nu\xi$. Thus in the limit of the one-dimensional hydrogen, $f(\xi)=1$ and we obtain again $\psi=xe^{-x/a}$, as we must.

We insert Eq.~(\ref{ansatzodd}) into Eq.~(\ref{dimlessSchrodinger}) and we obtain thus the differential equation that $f$ must satisfy:

\ba
\label{eqforfodd}
&&\frac{\xi}{2}(4\xi-1)f^{\prime\prime}(\xi)+(7\xi+\nu\xi-4\nu\xi^{2}-1)f^{\prime}(\xi)\nonumber\\
&&+(3-6\nu\xi-\frac{\nu^{2}\xi}{2}-\epsilon\nu^{2}\xi+2\nu^{2}\xi^{2})f(\xi)=0
\ea

Since $0\leq\xi\leq 1/4$, a series solution of this equation will converge rapidly. We can easily find then that

\be
\label{fexpansionodd}
f(\xi)=\sum_{i=0}^{\infty}j_{i}\xi^{i},
\ee

where

\ba
\label{eqj}
&&j_{0}=1\nonumber\\
&&j_{1}=3\nonumber\\
&&j_{2}=10-\nu-\frac{1}{6}(\nu^{2}+2\epsilon\nu^{2})\nonumber\\
&&j_{3}=35-\frac{31\nu}{6}-\frac{1}{18}(1+2\epsilon)\nu^{2}(15+\nu)\nonumber\\
&&j_{4}=\frac{1}{120}(15120-2652\nu-366\nu^{2}-840\epsilon\nu^{2}-20\nu^{3}\nonumber\\
&&-40\epsilon\nu^{3}-\nu^{4}+4\epsilon^{2}\nu^{4})\nonumber\\
&&j_{5}=\frac{1}{5400}(2494800-482940\nu-57834\nu^{2}-151200\epsilon\nu^{2}\nonumber\\
&&-2082\nu^{3}-4020\epsilon\nu^{3}-15\nu^{4}+600\epsilon\nu^{4}+1260\epsilon^{2}\nu^{4}\nonumber\\
&&-2\nu^{5}+40\epsilon\nu^{5}+88\epsilon^{2}\nu^{5})\nonumber\\
&&etc
\ea

Having found the exact series for $f(\xi)$, we can use Eq.~(\ref{normalization}) in order to find N. Then the energy functional of Eq.~(\ref{energy}) can be calculated as a function of the parameters $\nu$ and $\epsilon$. Requiring that the energy calculated be equal to the parameter $\epsilon$ would determine the value of $\epsilon$ in terms of $\nu$ for the various odd states. Each root $\epsilon$ of the resulting equation corresponds to an odd state. The lowest root corresponds to the first excited state.

There is an alternate way though to locate the value of $\epsilon$ that corresponds to the first excited state. We shall consider the energy functional for odd states that is calculated using Eq.~(\ref{fexpansionodd}) as a function of a variational parameter $\epsilon$ for a given value of $\nu$. Minimization of this function with respect to $\epsilon$ will yield the exact value of the lowest energy in terms of $\nu$, since the correct wavefunction of Eq.~(\ref{fexpansionodd}) has been used.

However, since the series is expected to converge rapidly, we can truncate the series at some point we choose. We can then determine N by putting the truncated wavefunction into Eq.~(\ref{normalization}) and use Eq.~(\ref{energy}) to calculate the energy. Our result will involve the parameters $\epsilon$ and $\nu$. However, if the full expression of Eq.~(\ref{fexpansionodd}) had been used our result would have been exact and the energy functional of Eq.~(\ref{energy}) would have its minimum at the correct value of $\epsilon$. Conversely, we can minimize the energy functional that arose through the use of the truncated $f(\xi)$, with respect to $\epsilon$. The function of $\nu$ we shall find will be an excellent approximation to the total energy.

We shall demonstrate below how rapidly this procedure converges and how we can obtain a very accurate expression for the energy of the first excited state as a function of the dimensionless distance $L/a=\nu$ between the protons.

A choice that is analogous to our ground state choice for a truncated $f(\xi)$ is $1+j_{1}\xi+j_{2}\xi^{2}$.

												Thus our wavefunction for this choice is

\be
\label{wavefunction2odd}
\psi(\xi)=n_{2}\xi\sqrt{1-4\xi}e^{-\nu\xi}(1+3\xi+j_{2}\xi^{2})
\ee
We can use Eq.~(\ref{normalization}) to find:

\be
\label{normalization2}
n_{2}=(\kappa_{0}+\kappa_{1}\epsilon+\kappa_{2}\epsilon^{2})^{-1/2},
\ee
where

\ba
\label{kappa0}
&&\kappa_{0}=\frac{3}{8192}+\frac{3378375}{1024\nu^{7}}+\frac{190575}{512\nu^{6}}+\frac{68985}{
4096\nu^{5}}\nonumber\\
&&+\frac{135}{64\nu^{4}}+\frac{4011}{16384\nu^{3}}-\frac{1051}{16384\nu^{2}}-\frac{631}{24576\nu}+\frac{31\nu}{147456}\nonumber\\
&&+\frac{\nu^{2}}{147456}+\frac{F(\sqrt{\nu/2})}{\sqrt{2\nu}}\bigl(\frac{163}{3072}-\frac{3378375}{512\nu^{7}}-\frac{1507275}{512\nu^{6}}\nonumber\\
&&-\frac{1177785}{2048\nu^5}-\frac{127995}{2048\nu^4}-\frac{33075}{8192\nu^3}-\frac{195}{4096\nu^2}+\frac{685}{8192\nu}\nonumber\\
&&-\frac{7\nu}{24576}-\frac{5\nu^2}{12288}-\frac{\nu^3}{73728}\bigr),
\ea

\ba
\label{kappa1}
&&\kappa_{1}=\frac{11}{3072}-\frac{225225}{1024\nu^5}-\frac{54285}{2048\nu^4}-\frac{9555}{4096\nu^3}-\frac{583}{4096\nu^2}\nonumber\\
&&+\frac{ 55}{6144\nu}+\frac{25\nu}{36864}+\frac{\nu^2}{36864}\nonumber\\
&&+\frac{F(\sqrt{\nu/2})}{\sqrt{2\nu}}\bigl(-\frac{23}{3072}+\frac{225225}{512\nu^5}+\frac{204435}{1024\nu^4}+\frac{85785}{2048\nu^3}\nonumber\\
&&+\frac{2625}{512\nu^2}+\frac{685}{2048\nu}-\frac{35\nu}{6144}-\frac{\nu^2}{768}-\frac{\nu^3}{18432}\bigr)
\ea

and
\ba
\label{kappa2}
&&\kappa_{2}=\frac{41}{6144}+\frac{15015}{4096\nu^3}+\frac{1925}{4096\nu^2}+\frac{133}{2048\nu}+\frac{19\nu}{36864}\nonumber\\
&&+\frac{\nu^2}{36864}+\frac{F(\sqrt{\nu/2})}{\sqrt{2\nu}}\bigl(-\frac{175}{1536}-\frac{15015}{2048\nu^3}-\frac{3465}{1024\nu^2}\nonumber\\
&&-\frac{1575}{2048\nu}-\frac{25\nu}{2048}-\frac{\nu^2}{1024}-\frac{\nu^3}{18432}\bigr).
\ea

We can also calculate the energy of Eq.~(\ref{energy}), to obtain:

\be
\label{energy2excited}
\epsilon_{2}=\frac{1}{\nu}+n_{2}^{2}(\gamma_{0}+\gamma_{1}\epsilon+\gamma_{2}\epsilon^{2})
\ee

where
\ba
\label{gamma0}
&&\gamma_{0}=-\frac{281}{294912}-\frac{7016625}{2048\nu^8}-\frac{1053675}{1024\nu^7}-\frac{593775}{8192\nu^6}-\frac{52035}{2048\nu^5}\nonumber\\
&&-\frac{129885}{32768\nu^4}-\frac{13439}{32768\nu^3}+\frac{7793}{49152\nu^2}+\frac{209}{49152\nu}-\frac{11\nu}{294912}\nonumber\\
&&+\frac{F(\sqrt{\nu/2})}{\sqrt{2\nu}}\bigl(-\frac{173}{16384}+\frac{7016625}{1024\nu^8}+\frac{4446225}{1024\nu^7}+\frac{4650975}{4096\nu^6}\nonumber\\
&&+\frac{840105}{4096\nu^5}+\frac{568485}{16384\nu^4}+\frac{45803}{8192\nu^3}+\frac{16121}{16384\nu^2}-\frac{481}{1536\nu}\nonumber\\
&&+\frac{15\nu}{8192}+\frac{11\nu^2}{147456}\bigr),
\ea
\ba
\label{gamma1}
&&\gamma_{1}=-\frac{215}{73728}+\frac{467775}{2048\nu^6}+\frac{302715}{4096\nu^5}+\frac{57645}{8192\nu^4}+\frac{8997}{8192\nu^3}\nonumber\\
&&+\frac{775}{12288\nu^2}+\frac{1}{2048\nu}-\frac{11\nu}{73728}\nonumber\\
&&+\frac{F(\sqrt{\nu/2})}{\sqrt{2\nu}}\bigl(-\frac{91}{12288}-\frac{467775}{1024\nu^6}-\frac{614565}{2048\nu^5}-\frac{342615}{4096\nu^4}\nonumber\\
&&-\frac{7385}{512\nu^3}-\frac{7095}{4096\nu^2}-\frac{859}{6144\nu}+\frac{17\nu}{3072}+\frac{11\nu^2}{36864}\bigr)
\ea
and

\ba
\label{gamma2}
&&\gamma_{2}=-\frac{149}{73728}-\frac{31185}{8192\nu^4}-\frac{10815}{8192\nu^3}-\frac{707}{4096\nu^2}-\frac{263}{12288\nu}\nonumber\\
&&-\frac{11\nu}{73728}+\frac{F(\sqrt{\nu/2})}{\sqrt{2\nu}}\bigl(\frac{469}{12288}+\frac{31185}{4096\nu^4}+\frac{10605}{2048\nu^3}\nonumber\\
&&+\frac{6405}{4096\nu^2}+\frac{895}{3072\nu}+\frac{23\nu}{6144}+\frac{11\nu^2}{36864}\bigr).
\ea

In all the above $F(x)=e^{-x^{2}}\int_{0}^{x}e^{y^{2}}dy$ is the Dawson integral.

The energy of Eq.~(\ref{energy2excited}) acquires its extrema at two values of $\epsilon$:

\ba
\label{plusminusexc}
&&\pm\frac{\sqrt{(\gamma_{2}\kappa_{0}-\gamma_{0}\kappa_{2})^{2}+(\gamma_{1}\kappa_{0}-\gamma_{0}\kappa_{1})(-\gamma_{2}\kappa_{1}+\gamma_{1}\kappa_{2})}}{\gamma_{1}\kappa_{2}-\gamma_{2}\kappa_{1}}\nonumber\\
&&+\frac{\gamma_{2}\kappa_{0}-\gamma_{0}\kappa_{2}}{\gamma_{1}\kappa_{2}-\gamma_{2}\kappa_{1}}.
\ea
However, we can easily ascertain that the upper sign corresponds to a maximum for all values of $\nu$. Thus we have to let

\ba
\label{extremumexc}
&&\epsilon=-\frac{\sqrt{(\gamma_{2}\kappa_{0}-\gamma_{0}\kappa_{2})^{2}+(\gamma_{1}\kappa_{0}-\gamma_{0}\kappa_{1})(-\gamma_{2}\kappa_{1}+\gamma_{1}\kappa_{2})}}{\gamma_{1}\kappa_{2}-\gamma_{2}\kappa_{1}}\nonumber\\
&&+\frac{\gamma_{2}\kappa_{0}-\gamma_{0}\kappa_{2}}{\gamma_{1}\kappa_{2}-\gamma_{2}\kappa_{1}}
\ea

If we insert this value of the variational parameter $\epsilon$ into the energy of Eq.~(\ref{energy2excited}), we obtain the final result for the energy:

\ba
\label{energy2finalexc}
&&\epsilon_{2}=\frac{\gamma_{1}\kappa_{1}-2\gamma_{0}\kappa_{2}-2\gamma_{2}\kappa_{0}}{\kappa_{1}^{2}-4\kappa_{0}\kappa_{2}}\nonumber\\
&&+\frac{2\sqrt{(\gamma_{2}\kappa_{0}-\gamma_{0}\kappa_{2})^{2}+(\gamma_{1}\kappa_{0}-\gamma_{0}\kappa_{1})(-\gamma_{2}\kappa_{1}+\gamma_{1}\kappa_{2})}}{\kappa_{1}^{2}-4\kappa_{0}\kappa_{2}}\nonumber\\
&&+\frac{1}{\nu}.
\ea

\begin{figure}[t]
\vskip 0.3cm
                        \includegraphics[width=0.49\textwidth]{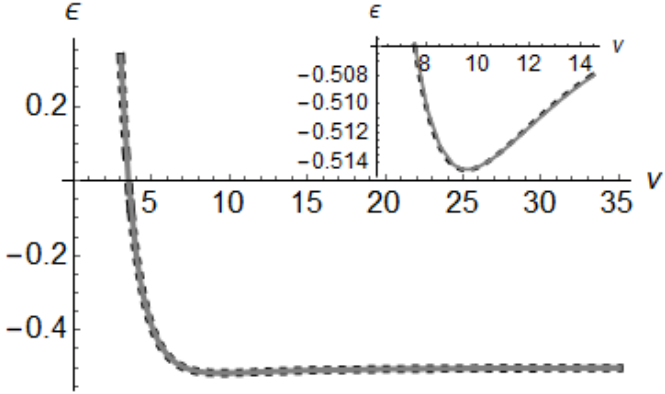}
                        \caption{\label{fig6} The dimensionless first excited state energy $\epsilon$ as a function of $\nu=L/a$ for the truncated wavefunction of Eq.~(\ref{wavefunction2odd}). The dashed curve is the analytical result of Eq.~(\ref{energy2finalexc}), while the continuous gray curve is the numerical result of Ref.\cite{Duan}. The inset shows the details of this comparison around the equilibrium point.}
                        \end{figure}

For small $\nu$ we find that $\epsilon_{2}\approx 3(4550-\sqrt{7728630})/(269\nu^{2})\approx 19.7393/\nu^{2}$, while for large $\nu$ $\epsilon_{2}\approx (-1/2)-3/(2\nu^{2})$. We note that $\epsilon_{2}$ acquires its minimum value of -0.514563 at $\nu=9.52343$. When $\nu=9.6$ it acquires the value -0.514557, while the corresponding numerical result is -0.514719\cite{Duan}. We clearly see in Figure~\ref{fig6} that the agreement of the analytical expression $\epsilon_{2}$ of Eq.~(\ref{energy2finalexc}) with the numerical results of Ref.\cite{Duan} is excellent for all values of $\nu$, even though the truncated function is very simple.

\begin{figure}[t]
\vskip 0.3cm
                        \includegraphics[width=0.49\textwidth]{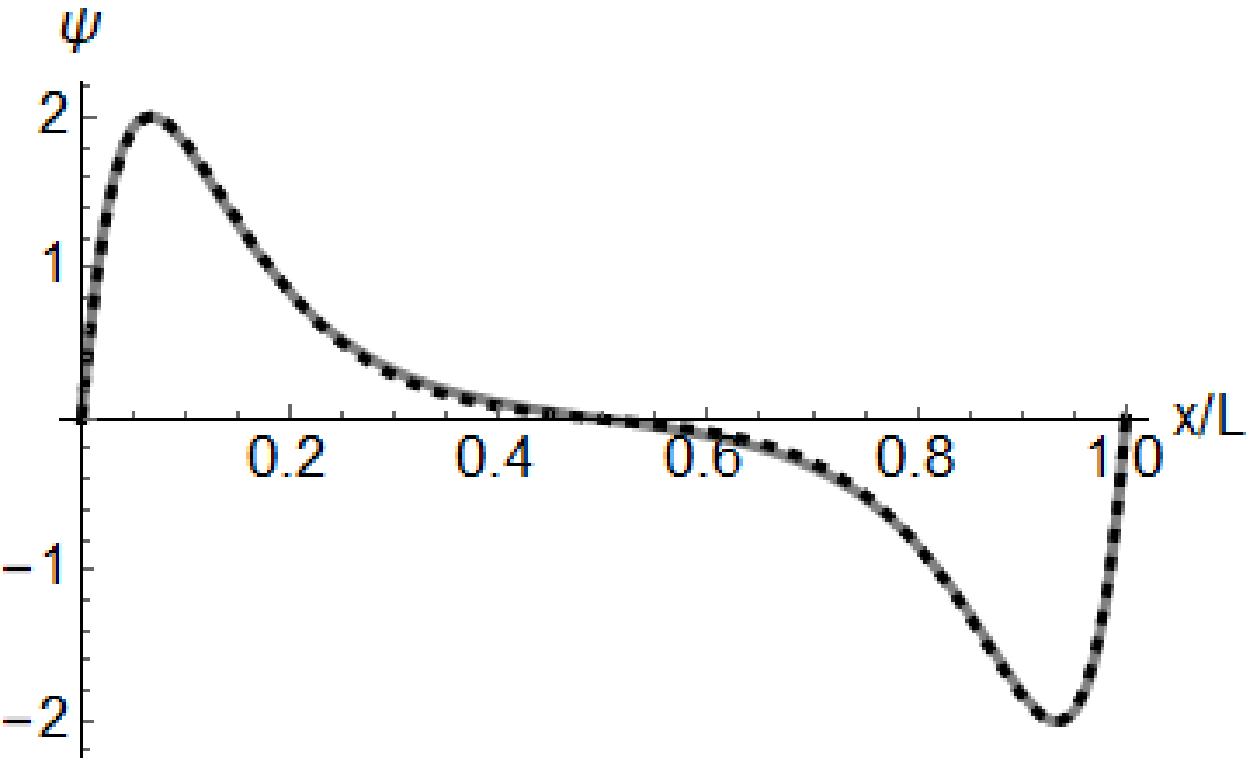}
                        \caption{\label{fig7}The first excited state wavefunction of Eq.~(\ref{wavefunction2odd}) (dashed curve) is compared to the corresponding wavefunction obtained numerically for the case $\nu=15$ (continuous gray curve).}
                        \end{figure}

		          		\begin{figure}[t]
\vskip 0.3cm
                        \includegraphics[width=0.49\textwidth]{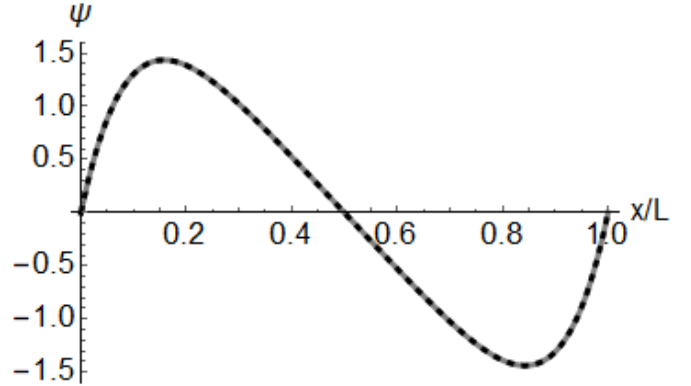}
                        \caption{\label{fig8}The first excited state wavefunction of Eq.~(\ref{wavefunction2odd}) (dashed curve) is compared to the corresponding wavefunction obtained numerically for the case $\nu=6$ (continuous gray curve).}
                        \end{figure}

We can also compare the wavefunction of Eq.~(\ref{wavefunction2odd}) with the wavefunction obtained numerically\cite{Alejandro}. This comparison is shown in Figure~\ref{fig7} for $\nu=15$ and in Figure~\ref{fig8} for $\nu=6$. We see again an excellent agreement between our analytical results and their numerical counterparts.

\begin{figure}[t]
\vskip 0.3cm
                        \includegraphics[width=0.49\textwidth]{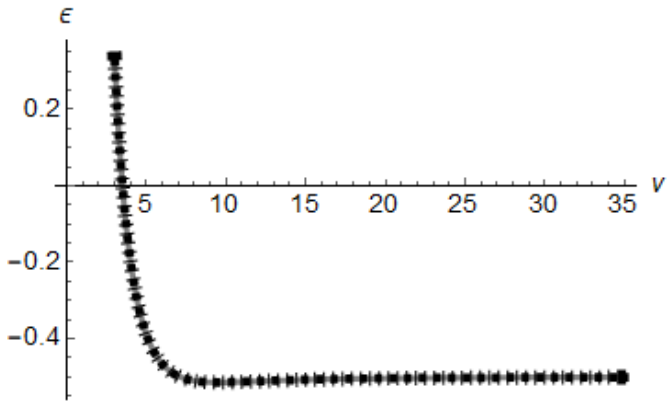}
                        \caption{\label{fig9}The dimensionless first excited state energy $\epsilon$ as a function of $\nu=L/a$ for the truncated wavefunction of Eq.~(\ref{wavefunction2odd}) and for truncated wavefunctions ending with the $j_{3}\xi^{3}$ and $j_{4}\xi^{4}$ terms. The dashed curve corresponds to the truncated wavefunction that ends with the $j_{2}\xi^{2}$ term, the continuous gray curve corresponds to the truncated wavefunction that ends with the $j_{3}\xi^{3}$ term, and the dotted curve corresponds to the truncated wavefunction that ends with the $j_{4}\xi^{4}$ term. All three curves coincide.}
                        \end{figure}

Actually, we can check the accuracy of our analytical results by examining the possibility of adopting as $f(\xi)$ the expression of Eq.~(\ref{wavefunction2odd}), augmented first by a $j_{3}\xi^{3}$ term and then by an additional $j_{4}\xi^{4}$ term. We find that the results obtained by using the actual Eq.~(\ref{wavefunction2odd}) are identical to the ones obtained when we use the augmentations mentioned above. This is most clearly seen in Figure~\ref{fig9}, where the dashed curve corresponds to the truncated wavefunction that ends with the $j_{2}\xi^{2}$ term, the continuous gray curve corresponds to the truncated wavefunction that ends with the $j_{3}\xi^{3}$ term, and the dotted curve corresponds to the truncated wavefunction that ends with the $j_{4}\xi^{4}$ term. All three curves coincide, showing thus that we need go no further than the wavefunction of Eq.~(\ref{wavefunction2odd}) in order to describe accurately the ground state of the one-dimensional molecular ion.									

\vskip 0.3cm
\vskip 0.3cm

{IV. \bf Conclusions}

We see that by condensing both boundary conditions into one, through the definition of the new variable $\xi$ that varies between 0 and 1/4, we can find an exact series solution in terms of $\xi$ that converges very rapidly. Then we can truncate the series, keeping terms up to $\xi^{2}$. We obtain thus a very accurate analytical closed form expression for the electronic energy as a function of the distance between the protons. This is the first time such a simple analytical solution has been found for the one-dimensional hydrogen molecular ion.

We note furhermore that this method can be generalized very easily to any one-dimensional problem that is defined on a finite interval and that requires the wavefunction to become zero at the ends of this interval. Then we can always rewrite the one-dimensional Schrodinger equation in terms of the variable $\xi$ defined by   Eq.~(\ref{xi}). Since $0\leq\xi\leq 1/4$, the one-dimensional Schrodinger equation can be solved by a rapidly converging series of powers of $\xi$. Keeping just the first few terms of this series will give an extremely accurate analytical solution.
We can demonstrate the value of this method by finding the $\xi$ solution for the ground state of the one-dimensional box. We can easily show that the simplest resulting solution is proportional to $\xi+\xi^2$ and gives an energy of $153\hbar^{2}/(31mL^{2})$. The agreement of both of these approximations to the corresponding exact results is excellent.

\end{document}